\documentclass[12pt,preprint]{aastex}
\usepackage{emulateapj5}

\newcommand{\lsim}{\raisebox{-0.3ex}{\mbox{$\stackrel{<}{_\sim} \,$}}}

\def\gta{\ifmmode {\mathbin{\lower 3pt\hbox   
    {$\,\rlap{\raise 5pt\hbox{$\char'076$}}\mathchar"7218\,$}}}
    \else {${\mathbin{\lower 3pt\hbox
    {$\rlap{\raise 5pt\hbox{$\char'076$}}\mathchar"7218\,$}}}
    $}\fi}
\def\lta{\ifmmode {\,\mathbin{\lower 3pt\hbox   
    {$\,\rlap{\raise 5pt\hbox{$\char'074$}}\mathchar"7218\,$}}}
    \else {${\mathbin{\lower 3pt\hbox
    {$\rlap{\raise 5pt\hbox{$\char'074$}}\mathchar"7218\,$}}}
    $}\fi}

\shorttitle {An unusual precursor burst with oscillations from SAX J1808.4--3658} \shortauthors {Bhattacharyya \& Strohmayer}

\begin{document}

\title{An unusual precursor burst with oscillations from SAX
J1808.4--3658}

\author {Sudip Bhattacharyya\altaffilmark{1,2}, and Tod
E. Strohmayer\altaffilmark{1}}

\altaffiltext{1}{X-ray Astrophysics Lab,
Exploration of the Universe Division,
NASA's Goddard Space Flight Center,
Greenbelt, MD 20771; sudip@milkyway.gsfc.nasa.gov,
stroh@clarence.gsfc.nasa.gov}

\altaffiltext{2}{Department of Astronomy, University of Maryland at
College Park, College Park, MD 20742-2421}

\begin{abstract}

We report the finding of an unusual, weak precursor to a thermonuclear
X--ray burst from the accreting millisecond pulsar SAX J1808.4--3658.
The burst in question was observed on Oct. 19, 2002 with the Rossi
X-Ray Timing Explorer (RXTE) proportional counter array (PCA). The
precursor began $\approx 1$ s prior to the onset of a strong radius
expansion burst, lasted for about 0.4 s, and exhibited strong
oscillations at the 401 Hz spin frequency. Oscillations are not
detected in the $\approx 0.5$ s interval between the precursor and the
main burst. The estimated peak photon flux and energy fluence of the precursor
are about 1/25, and 1/500 that of the main burst, respectively. From joint
spectral and temporal modeling, we find that an expanding burning
region with a relatively low temperature on the spinning neutron star
surface can explain the oscillations, as well as the faintness of the
precursor with respect to the main part of the burst.  We dicuss some
of the implications of our findings for the ignition and spreading of
thermonuclear flames on neutron stars.

\end{abstract}


\section {Introduction} \label{sec: 1}

Four strong thermonuclear (type-I) X--ray bursts were observed with
the {\it RXTE} proportional counter array (PCA) from the accreting
millisecond pulsar SAX J1808.4--3658 when this source was in outburst
in 2002 (Chakrabarty et al. 2003).  Such bursts are produced by
thermonuclear burning of matter accumulated on the surfaces of
accreting neutron stars (Woosley, \& Taam 1976; Lamb, \& Lamb 1978).
All the bursts exhibited strong brightness oscillations near the known
stellar spin frequency ($\approx 401$ Hz; Wijnands \& van der Klis
1998), which confirmed that this timing feature originates at the
neutron star surface. During the burst rise, an expanding burning
region (hot spot) on the spinning stellar surface may give rise to
these oscillations (Strohmayer, Zhang \& Swank 1997; Miller, \& Lamb
1998; Nath, Strohmayer, \& Swank 2002), while during the burst decay
(when the whole stellar surface may be engulfed by thermonuclear
flames), the origin of this timing feature may be temperature
variations due to surface waves (Heyl 2005; Lee \& Strohmayer 2005;
Cumming 2005; Piro \& Bildsten 2006). 
Three of these bursts (Oct 15, 18 and 19) from SAX
J1808.4--3658 exhibited strong oscillations during the intensity rise.
Bhattacharyya \& Strohmayer (2006c) found evidence for complex
variation of the oscillation frequency during the rise of the Oct. 15
and 18 bursts. The Oct. 19 burst did not show evidence for similar
variation, although the other properties of this burst were akin to
those of the Oct. 15 and 18 bursts.

Analysis of high time resolution lightcurves just prior to the bursts
reveal a weak precursor event to the Oct. 19 burst. To our knowledge
this is the first report of such a precursor to a normal, hydrogen --
helium powered thermonuclear burst. Several superbursts, which are
likely powered by fusion of heavier elements (Strohmayer \& Brown
2002; Cumming \& Bildsten 2001; Schatz et al. 2002), have shown
precursor events which have the characteristics of shorter, normal
bursts.  The precursor to the Oct. 19 burst looks like a typical
thermonuclear burst except it lasts less than a second, and has a peak
photon flux only 1/25 of the main burst. Also unique is the fact that
strong pulsations are detected during this precursor.  The Oct. 15 and
18 bursts do not show a similar precursor.  In this Letter we describe
the properties of the precursor, and discuss the implications for its
size and oscillation content in the context of ignition and spreading
of thermonuclear instabilities on neutron stars.

\section {Data Analysis and Results} \label{sec: 2}

We analyzed the Oct. 19, 2002 archival {\it RXTE}-PCA data from SAX
J1808.4--3658. During this observation the source was in outburst, and
the data contain a thermonuclear X--ray burst, some of the properties
of which were reported in Chakrabarty et al. (2003). We found an
excess of intensity less than a second prior to the rise of this
burst, that lasted for $\le 0.5$ s. Figure 1 shows the lightcurve of
the burst at 1/16 s using a logarithmic intensity scale. The precursor
is evident as the spike just prior to the rising edge of the main part
of the burst. The precursor has a peak count rate of 2300 s$^{-1}$
(1/16 s intervals, 4 PCUs), while the average persistent count rate
prior to this feature was $\sim 640$ s$^{-1}$. This shows that the
feature is significant. The rapid rise and slower decay of intensity
during the precursor (see Fig. 2) is similar to that seen in most
normal bursts, only the peak intensity and timescale are smaller and
shorter, respectively. The spectra of thermonuclear bursts can usually
be modelled with a blackbody function (Strohmayer \& Bildsten
2006). We, therefore, fitted the precursor spectrum (using 0.25 s of
data) with an absorbed blackbody and found a temperature of
$1.26^{+0.10}_{-0.13}$ keV (reduced $\chi^2 \approx 1.4$, 28 degrees
of freedom, see Figure 2). The reduced $\chi^2$ is acceptable, and
supports the idea that the precursor is indeed thermonuclear in
origin.

Next, we searched for oscillations in the 0.3 s of event mode data
(shown with dotted lines in Figure 2) for which the precursor
intensity was significantly above the persistent level. We computed a
power spectrum with a Nyquist frequency of $2048$ Hz, and a frequency
resolution of $3.3$ Hz, and found a peak (power level $\approx 43.9$)
near the known stellar spin frequency $\sim 401$~Hz (Fig. 3). As burst
oscillation frequencies are not known to evolve by more than $\approx
6$ Hz (Giles et al. 2002; Muno et al. 2002; Bhattacharyya \&
Strohmayer 2005), considering the number of trials $N_{\rm trial} =
2$, we have a significance of $6.01 \times 10^{-10}$, which implies a
$> 6\sigma$ detection of oscillations during the precursor. The
fractional rms amplitude for this 0.3 s interval is $A_{\rm 1} =
0.375\pm0.068$ (reduced $\chi^2 = 7.0/13$ from fitting a
constant+sinusoid model to the persistent emission subtracted,
phase-folded light curve), and no significant harmonic component was
detected. This high amplitude and the comparatively broad peak in the
power spectrum show that these oscillations originate from the
precursor, and are not accretion-powered pulsations. We also
calculated dynamic $Z^2$ power spectra (Strohmayer \& Markwardt
1999), and the corresponding power contours. Figure 2 shows that 
there are two sets of disconnected power 
contours, one during the precursor, and another during the rise of
the main burst. This shows that the oscillations
during the precursor are unique to it, and, are not connected to the 
main burst.

Now, several important questions are: (1) why does the precursor occur
about a second prior to the main burst, and (2) why was it so much
fainter than the main burst.  Note that this burst may not be
considered as a single double-peaked non-photospheric radius expansion
(non-PRE) burst, because such bursts are always weak
(ie. sub-Eddington, Sztajno et al. 1985), while the main burst was a
strong PRE burst. Therefore, sequential emissions from two different
portions of the neutron star surface (Bhattacharyya \& Strohmayer
2006a; 2006b), or a two step energy generation due to convective
mixing of the nuclear fuel (Fujimoto et al. 1988) probably cannot
explain the precursor event. A two step energy release might answer
the first question. The second question may be addressed in the
following way. Near the peak of the main burst, most of the neutron
star surface is expected to emit near the Eddington
temperature. Therefore, a smaller emission region and/or a lower
temperature during the precursor would explain its faintness compared
to the main burst.  This might happen in three ways: (1) if the fuel
for the precursor is confined to a small portion of the stellar
surface (possibly by the magnetic field) and the burning region has a
high temperature (the small hot spot can produce high amplitude, spin
modulated pulsations); (2) if the thermonuclear flame spreads all over
the stellar surface in $ < 0.1$ s at the onset of the precursor, and
the whole surface emits at a low temperature (in such a case, surface
modes might account for the oscillations; see \S~1); and (3) if the
thermonuclear flame with an intermediate average temperature takes
$\sim 0.2-0.3$ s to spread (in this case, the expanding hot spot may give
rise to the oscillations). Joint timing and spectral modeling can help
us discriminate between these alternatives.

To be more specific about the joint analysis, 
let us first assume that the observed photon flux (after the subtraction
of the persistent emission) in any time bin during a burst is $F$. Then 
we note that the ratio ($Ratio$) of $F$
near the peak of the main burst (the last time bin of the lower
panel of Fig. 2; hereafter bin 1) to $F$ during the precursor (the
first bin of the lower panel of Fig. 2; hereafter bin 2) is
$25.1\pm1.5$ (the energy flux ratio is about 45 to 1). The corresponding
observed blackbody temperatures during these bins are $T_{\rm obs,1} =
2.94\pm0.13$ keV and $T_{\rm obs,2} = 1.26^{+0.10}_{-0.13}$ keV
(mentioned before) respectively. We also note that the fractional rms
amplitude of oscillations during bin 2 is $ A_{\rm 2} =
0.403\pm0.071$. From our spectral modeling we also find that the
precursor contained only $\approx 1/400$ of the energy in the main
burst. Now for the joint modeling, assuming the stellar and other
source parameter values, one needs to reproduce the oscillation
amplitude $ A_{\rm 2}$, and then, with the same parameter values
(including the average burning region size), one needs to reproduce
$Ratio$ from $T_{\rm obs,1}$ and $T_{\rm obs,2}$. We do this in
\S~3.2.

\section {Comparison with Models}

The primary aim of our modeling is to understand both the faintness of
the precursor (relative to the peak of the burst), and the presence of
high amplitude brightness oscillations.  In our simple model, we
assume emission from a circular burning region (hot spot) on the
rotating stellar surface (Bhattacharyya et al. 2005; see also some other
works, e.g., Muno, \"Ozel, \& Chakrabarty 2002; Miller \& Lamb 1998;
Cadeau et al. 2006; Braje, Romani, \& Rauch 2000; Poutanen \&
Gierli\'nski 2004).  Brightness
oscillations occur as the image of the hot spot in the observer's sky
periodically changes with the stellar spin. The corresponding
fractional rms amplitude ($A$) can be determined by fitting the
phase-folded light curve (normalised to have the observed count
rate). The total observed photon flux can also be computed from the
blackbody spectrum for an assumed temperature. For these calculations,
we combine the model with the appropriate instrument response matrix.
Our model includes the following physical effects: (1) Doppler effect
due to rapid stellar spin, (2) special relativistic beaming, and (3)
gravitational redshift and light bending (assuming Schwarzschild
spacetime). In our numerical calculations, we track the paths of
photons in order to incorporate the light bending effect in the
calculated photon flux (Bhattacharyya, Bhattacharya, \& Thampan
2001). We use the following parameters in our model: (1) neutron star
mass $M$ (in $M_\odot$), (2) dimensionless stellar radius-to-mass
ratio $Rc^2/GM$, (3) stellar spin frequency $\nu$ ($\approx 401$ Hz;
\S~1 ), (4) observer's inclination angle $i$ measured from the upper
rotational pole, (5) polar angle of the hot spot center $\theta_{\rm
c}$, (6) angular radius of the hot spot $\Delta\theta$, and (7) the
blackbody temperature $T_{\rm BB}$. For SAX J1808.4--3658, Li et
al. (1999) and Bhattacharyya (2001) calculated constraints on $M$ and
$Rc^2/GM$ (although they assumed that the stellar magnetic field is
entirely dipolar).  For example, if the lower limit of $Rc^2/GM$ is
4.0, the upper limit of $M$ is $\sim 1.4$ (equation 6 of Bhattacharyya
2001). Here for our illustrative model, we assume $M=1.4$ and
$Rc^2/GM=4.0$ (i.e., radius $R \approx 8.3$ km). 
However, other values of $M$ and $Rc^2/GM$ in
reasonable ranges do not alter our conclusions significantly.  In our
calculations, we mostly use $i = 60^{\rm o}$, as this is the average
value for a randomly oriented stellar spin axis. We use $i \approx
80^{\rm o}$ as the upper limit, because the absence of a deep eclipse
indicates $i < 82^{\rm o}$ (Chakrabarty \& Morgan 1998). For a source
distance $d = 3$ kpc, Wang et al. (2001) suggested $i \approx 20^{\rm
o}-65^{\rm o}$ (with 90\% confidence) based on the modeling of the
X--ray and optical emission from SAX J1808.4--3658. A higher value of
$d$ ($3.4-3.6$ kpc; Galloway \& Cumming 2006) would shift this range
of $i$ towards slightly smaller values.  We, therefore, consider the
case $i = 40^{\rm o}$ in our joint spectral and timing modeling in
\S~3.2. We vary the other parameter values for our model calculations.

\subsection {Inferences from Timing Data}

Before conducting the joint spectral and timing modeling, we explore
whether or not a hot spot model can reproduce the observed amplitude
$A_1 = 0.375\pm0.068$ (\S~2) during the precursor.  A burning region
with $i=60^{\rm o}$, $\theta_{\rm c}=60^{\rm o}$,
$\Delta\theta=60^{\rm o}$ and the observed $T_{\rm BB}=1.26$ keV gives
the amplitude $A = 0.364\pm0.048$, which is consistent with $A_1$.
Here we note that the harmonic content of this model light curve
cannot be significantly detected because of the small number of
observed counts, consistent with the observations. Next, we change
$T_{\rm BB}$ to $=1.0$ keV, which does not alter $A$ ($=
0.372\pm0.048$) much, and $i=80^{\rm o}$ can also reproduce $A_1$ for
similar values of $\theta_{\rm c}$, $\Delta\theta$ and $T_{\rm
BB}$. We note that $A$ increases with the increase of $i$ and
$\theta_{\rm c}$ (up to $\theta_{\rm c} = 90^{\rm o}$), and with the
decrease of $\Delta\theta$. For example, keeping $i=60^{\rm o}$ and
$T_{\rm BB}=1.26$ keV, if we change $\theta_{\rm c}$ to $45^{\rm o}$,
then to reproduce an $A$ ($= 0.354\pm0.048$) which is consistent with
$A_1$, $\Delta\theta$ has to be $5^{\rm o}$. This demonstrates two
points: (1) for fixed values of other parameters and a lower limit on
$\Delta\theta$ (such a lower limit should exist, as the burning region
has to have a finite size; Spitkovsky, Levin, \& Ushomirsky 2002),
there is a lower limit on $\theta_{\rm c}$; and (2) the size of the
burning region cannot be meaningfully constrained from the lower side
using the observed oscillation amplitude alone. However, this size can
be constrained from the upper side, as $i=80^{\rm o}$ and $\theta_{\rm
c}=90^{\rm o}$ (that allow the near-maximum value of $\Delta\theta$
for a given oscillation amplitude $A_1$) gives $\Delta\theta_{\rm max}
\sim 90^{\rm o}$. Therefore, our modeling of the timing data shows
that a hot spot, that does {\it not} encompass most of the stellar
surface, can give rise to the observed oscillation amplitude.

\subsection {Joint Spectral and Timing Inferences: An Illustration}

For our joint modeling (last paragraph of \S~2), we assume emission
from the whole stellar surface during bin 1 (a time bin during the
peak of the main burst; see \S~2).  This is because a smaller burning
region would give rise to significant oscillations (which are not
observed at this time), and would probably imply a super-Eddington
luminosity.  For this analysis, we consider the same values of $M$,
$Rc^2/GM$, $\nu$ and $i$ (as mentioned earlier in this section), and
vary the values of $\theta_{\rm c}$ and $\Delta\theta$ for the
precursor. However, for $T_{\rm BB}$, we use the surface color
temperature $T_{\rm c}$, which is related to the observed temperature
$T_{\rm obs}$ by $T_{\rm c} = T_{\rm obs}(1+z)$ (where the surface
gravitational redshift $1+z = (1-2GM/Rc^2)^{-1/2}$).  Moreover, due to
spectral hardening in the neutron star atmosphere, the effective
surface temperature ($T_{\rm eff}$) is related to $T_{\rm c}$ by
$T_{\rm eff} = T_{\rm c}/f$, where the color factor $f$ is greater
than 1 (London, Taam, \& Howard 1984). Therefore, in order to
calculate the observed photon flux, we use $(1/f^4)B(E_{\rm em},T_{\rm
c})$ as the emitted specific intensity (Fu \& Taam 1990; Bhattacharyya
et al. 2001). Here, $B$ is the Planck function and $E_{\rm em}$ is the
energy of a photon in the emitter's frame.  \"Ozel (2006) has recently
suggested the following expression for $f$ (based on the model
atmosphere calculations of Madej, Joss, \& R\'o\.za\'nska 2004): $f =
1.34+0.25((1+X)/1.7)^{2.2}((T_{\rm eff}/10^7{\rm
K})^4/(g/10^{13}{\rm cm}/{\rm s}^2))^{2.2}$.  Here, the surface gravitational
acceleration $g$ is given by $(GM/R^2)(1-2GM/Rc^2)^{-1/2}$, and $X$ is
the hydrogen mass fraction. For our illustrative model, initially we
assume the cosmic abundance $X=0.7$ for both bin 1 and bin 2 (time
bins from Fig. 2; see the last paragraph of \S~2). However, we note
that a change in the value of $X$ does not alter our timing results,
as the oscillation amplitude does not depend on $f$, and hence on $X$.
For our assumed values of the parameters, $f = 1.805$ ($T_{\rm eff,1}
= 2.30$ keV; bin 1) and $1.344$ ($T_{\rm eff,2} = 1.33$ keV; bin 2).
Here we note that, although $T_{\rm eff,1}$ is high, the corresponding
luminosity is less than (but close to) the Eddington luminosity, which
shows the consistency among our assumed parameter values.

Now, we follow the procedure that is described in the last paragraph
of \S~2.  First, we calculate the photon flux (for $f=1.805$) for the
emission from the whole stellar surface at the color temperature
$T_{\rm c} = 4.16$ keV (corresponding to $T_{\rm obs,1}$; bin 1).
Then we compute the oscillation amplitude $A$ and the photon flux for
bin 2 using $T_{\rm BB}=1.78$ keV (i.e., $T_{\rm c}$ corresponding to
$T_{\rm obs,2}$), $f = 1.344$, and various values of $i$, $\theta_{\rm
c}$ and $\Delta\theta$.  For $i = 60^{\rm o}$, $\theta_{\rm c}=65^{\rm
o}$ and $\Delta\theta = 55^{\rm o}$, we find $A = 0.398\pm0.050$ and
$Ratio = 25.3$, which are consistent with $A_2$ and the observed value
of $Ratio$ respectively. These two observed parameter values can also
be reproduced for $i = 80^{\rm o}$ for slightly different values of
$\theta_{\rm c}$ and $\Delta\theta$.  Moreover, for $i = 40^{\rm o}$,
$\theta_{\rm c}=110^{\rm o}$ and $\Delta\theta = 67^{\rm o}$, we get
$A = 0.391\pm0.050$ and $Ratio = 25.7$.  These show that our simple
hot spot model is consistent with the timing and the spectral data
simultaneously, and the inferred size of the hot spot (i.e., burning
region) is similar to that inferred in \S~3.1. For $i = 60^{\rm o}$,
$\theta_{\rm c}=50^{\rm o}$, and $\Delta\theta =5^{\rm o}$, $A$
($=0.378\pm0.050$) is consistent with $A_2$, but $Ratio$ ($=2504.3$)
is widely different from the observed value.  This shows that the
spectral data do not allow a small hot spot for the precursor
burst. In fact, for a variable $\theta_{\rm c}$ and for $i = 60^{\rm
o}$ (and other assumed parameter values), $\Delta\theta$ cannot be
much less than $55^{\rm o}$.  As in \S~3.1, we next try to determine
$\Delta\theta_{\rm max}$ for $i = 80^{\rm o}$ and $\theta_{\rm
c}=90^{\rm o}$. We can reproduce $A_2$ well for $\Delta\theta =
85^{\rm o}$, but the corresponding $Ratio$ ($=13.8$) is much less than
the observed value. Therefore, spectral data indicate that
$\Delta\theta < 85^{\rm o}$, and support the inference from the timing
data (\S~3.1), that the average angular radius of the burning region
of the precursor cannot be much larger than $90^{\rm o}$. In Fig. 4, 
we summarize these results.

In the previous paragraph, we assumed $X=0.7$ for both the time
bins. But if the precursor (bin 2) and the main burst (bin 1) were
ignited at different layers of accreted matter (\S~2), $X_{\rm prec}$
might be greater than $X_{\rm main}$. Here we assume the extreme
values ($X_{\rm prec} = 0.7$, i.e., hydrogen-rich, and $X_{\rm main} =
0.0$, i.e., helium-rich), and check if the inferences of the previous
paragraph still hold. Clearly, the new value of $X_{\rm main}$ alters
(increases) only the photon flux for bin 1 (as $f$ becomes $1.654$),
and hence the value of $Ratio$ changes. For $i = 60^{\rm o}$,
$\theta_{\rm c}=75^{\rm o}$ and $\Delta\theta = 70^{\rm o}$ (for bin
2), we find $A = 0.389\pm0.050$ and $Ratio = 25.3$, which are
consistent with the observed values. Therefore, our simple hot spot
model can simultaneously explain both the timing and the spectral
data, even when the chemical composition of the burning matter of the
two bursts are very different.  For $X_{\rm main} = 0.0$, as the value
of $f$ (for bin 1) decreases, and hence the corresponding model photon
flux increases, the model photon flux (and hence the hot spot size) of
the precursor has to increase in order to reproduce the observed
$Ratio$.  Therefore, a small hot spot is even more disfavored for
$X_{\rm prec} = 0.7$ and $X_{\rm main} = 0.0$. But do these extreme
values of $X$ allow a precursor burning region that is much larger
than $90^{\rm o}$?  For $i = 80^{\rm o}$ and $\theta_{\rm c}=90^{\rm
o}$ (for bin 2), we can reproduce $A_2$ for $\Delta\theta = 85^{\rm
o}$, but the corresponding $Ratio$ ($=19.7$) is significantly less than the
observed value. Therefore, even when the chemical composition of the
bursts are considerably different, the modeling of the spectral data
indicate $\Delta\theta \lsim 90^{\rm o}$ for the precursor.

The results of our modeling show that the hot spot during the
precursor burst was neither small, nor large enough to cover most of
the stellar surface.  This, along with the faintness of the precursor
compared to the main burst, argues against the scenarios 1 and 2 (mentioned
in \S~2).
The scenario 2 is further disfavored, because the large 
observed oscillation amplitude cannot originate from surface modes.
Therefore, as the joint analysis suggests, the oscillations were
produced by a hot spot (burning region) of moderate size 
($\Delta\theta \sim 50^{\rm o}-75^{\rm o}$). Now, if the 
burning region expansion does not happen, then the fuel
(accreted matter) has to be confined in this hot spot. 
But it is very difficult to understand what confines the fuel.
Magnetic field cannot do it, because, (1) the polar cap is unlikely to be
as big as the hot spot, (2) the magnetic field of SAX J1808.4--3658 may be
of the order of $10^8$ to $10^9$ G (Psaltis \& Chakrabarty 1999),
which cannot confine the fuel, and (3) Bhattacharyya \& Strohmayer (2006c)
have argued that thermonuclear flame
spreading likely occurs in the case of SAX J1808.4-3658
(and hence the fuel does not remain confined).
Therefore, it is very likely that before the precursor, there was 
accreted matter all over the stellar surface, and during the precursor, 
the burning region (with a relatively low temperature) expanded for $\sim
0.2-0.3$ s to engulf the whole stellar surface (scenario 3 of \S~2). 
In such a case, the $\Delta\theta$ of our model represents an 
average angular radius (during time bin 2), which is consistent with
the moderate hot spot size found from the joint analysis.
The expanding burning region can naturally account
for the observed oscillations, and when the burning covered most of
the stellar surface, the oscillations ceased.

\section {Discussion and Conclusions}

In this Letter, we have reported the discovery of a unique precursor
to a thermonuclear burst that; (1) occured about a second prior to the
main burst, (2) existed for a portion of a second, (3) had a peak
intensity more than a order of magnitude less than that of the main
peak, and (4) showed strong spin modulation pulsations. With
relatively simple modeling, we have found that an expanding burning
region at a relatively low temperature can explain the oscillations,
as well as the faintness of the precursor.

The low temperature and fluence of the precursor (compared to the main
peak) suggests that the amount of fuel involved in the precursor is
small compared to the total available for the whole burst.  It would
seem that there are at least three possibilities to account for the
precursor.  In the first scenario the release of nuclear energy at
depth could have such a two-step time dependence.  In this case the
observed time dependence would be a direct reflection of the time
dependent energy release due to nuclear burning. In the second class
of models, the precursor could be produced by a physical separation of
fuel layers. This, combined with the finite energy transport
time-scale through the surface layers results in the observed two-step
energy release.  This scenario, or one very like it, is thought to be
responsible for the precursors observed with superbursts (see
Strohmayer \& Brown 2002; Strohmayer \& Markwardt 2003; Weinberg et al.
2006).  In
superbursts, ignition is thought to occur via unstable carbon burning,
perhaps in a background of heavy {\it rp}-process ashes (Cumming \&
Bildsten 2001; Schatz et al. 2001; Strohmayer \& Brown 2002). This
occurs at much greater column depths ($\approx 2\times 10^{12}$ g
cm$^{-2}$; Cumming et al. 2006) 
than the unstable helium burning which ignites normal
bursts ($\approx 2\times 10^8$ g cm$^{-2}$; Woosley et al. 2004).  The
energy released at depth by a superburst diffuses upwards and,
partially, inwards. The outward going flux triggers the hydrogen -
helium fuel above it, resulting in the precursor burst.  In this case,
the combination of radial separation of the fuel layers and finite
energy diffusion time-scales results in the observed precursor.

A third possibility is that the precursor acts as a ``trigger'' which
initiates the burst.  If unstable burning begins somewhere on the star
at a column depth above that where simple considerations of the
ignition physics would suggest then it could act as a ``spark,''
setting off the remaining combustible fuel below.  To date, most
theoretical investigations have considered ignition conditions based
on spherically symmetric perturbations. Recent observations, in the
context of burst oscillations, suggest that non-axisymmetric processes
are likely crucial for a complete understanding of ignition and
spreading.  Recent theoretical work has also reached this conclusion
(see Spitkovsky, Levin \& Ushomirsky 2002). Perhaps temperature,
composition, and or accretion gradients across the stellar surface
might bring about such a condition.  Once nuclear energy release
begins locally, then heat will flow from that layer both in and out.
If ignition conditions are relatively finely ``balanced,'' then it
might not take much additional heat flux to set off the rest of the
fuel.  While the physical quantities which govern ignition are almost
certainly not uniform across the star, it remains uncertain whether
such conditions vary enough to make this kind of triggering possible.

Can we say whether either of these alternatives is at work (or not) in
the October 19, 2000 burst?  Based on the recent study of Galloway \&
Cumming (2006) it seems highly likely that the bursts from SAX
J1808.4--3658 were
ignited in a helium rich environment. While the exact nuclear
composition is not known, and indeed, details of some of the relevant
nuclear processes are uncertain, it seems unlikely, though not
impossible, that a pure helium ignition would have such a delayed
energy release. For example, the calculations of Woosley et al. (2004)
indicate a rapid, monotonic rise of the luminosity (without any precursor) 
from helium-rich ignitions, 
although we note that these were one-dimensional (radial)
calculations.  We suggest it is more likely that a situation like the
2nd or 3rd scenarios is responsible for the precursor, but it is
difficult to be more precise with only the one example at present.

An additional clue might be the fact that the Oct.  19 burst was the
last observed (of the 4 bursts detected), and happened at the lowest
accretion rate (see Galloway \& Cumming 2006).  Previous theoretical
work has shown that there exists an accretion rate regime in which
bursts can be triggered by unstable hydrogen burning (Fujimoto Hanawa
\& Miyaji 1981; Fushiki \& Lamb 1987; Narayan \& Heyl 2003). Galloway
\& Cumming (2006) argue that the bursts observed from SAX J1808.4--3658 in October
2002 were in or near the pure helium shell ignition regime, with the 
accretion rate per unit area of the stellar surface $\dot
m \sim 1,000$ g cm$^{-2}$ s$^{-1}$.  This is close to the critical $\dot m$
below which stable hydrogen burning switches off (see, for example,
Fujimoto et al. 1981).  If $\dot m$ dropped below this threshold some
time after the 3rd burst, then an accumulating hydrogen layer could
form above a partially formed helium layer.  If unstable burning were
then triggered in the hydrogen layer, it would inject enough energy to
raise the temperature and stabilize the hydrogen burning, but it might
not take much additional energy input to destabilize the helium layer
below, and set off the remainder of the accreted fuel.  This could
explain the faintness of the precursor in that the burning timescale
is longer for the temperature-dependent CNO cycle (than for the
triple-$\alpha$ process), and/or the fact that it might only take a
small amount of energy to stabilize the hydrogen burning once started
(Fujimoto et al. 1981) .  We speculate that such a process might
explain the precursor.

If this reasoning is correct then the duration of the precursor,
$t_{dur}$, and the time between it's start and the rising edge of the
main burst, $\Delta t_{pre} \approx 1$ s, can provide some rough
constraints on several relevant time-scales.  If energy must flow
inward to trigger the main burst, and then flow back out--for us to
see the main burst--then $\Delta t_{pre}$ is approximately twice the
radiative diffusion time from the trigger layer to the column depth of
helium fuel responsible for the main burst.  This gives a diffusion
time-scale of $\approx 0.5$ s, which is roughly consistent with
theoretical calculations (Cumming \& Bildsten 2000).  While we do not
claim to be able to precisely infer this quantity, the fact that it is
in qualitative agreement with theoretical expectations provides some
support for the idea of radially separated fuel layers.

What about spreading time-scales?  For this scenario to work the
intensity profile of the precursor would have to be largely controlled
by the spreading of the hydrogen burning layer.  In order to account
for the overall rise-time and duration of the precursor, the spreading
time would have to be approximately several tenths of seconds.
Indeed, the rise and decay of the precursor would directly represent
lateral spreading across the surface. This time-scale would also
accomodate the observed duration of the oscillations during the
precursor. One would likely require that the cooling time for the
hydrogen layer be less than or of order the spreading time, or else
the decay of the precursor would be difficult to understand. We note
that there is evidence for weak emission between the precursor and
main burst (see Figure 1), which provides some support for the idea
that the initial (possibly hydrogen) energy release had a longer
time-scale.  Finally, we reported evidence that this burst had a
somewhat different frequency evolution of the oscillations observed on
the rising edge (of the main burst) compared to the two other bursts
(Bhattacharyya \& Strohmayer 2006c).  It may be possible that the
different ignition condition suggested here also contributed to this
difference.

While the arguments above seem to provide a reasonable qualitative
description, they will remain largely speculative until more detailed
calculations are done.  It is another indication that high quality
data is forcing us to explore interesting details of nuclear burning
on neutron stars. Indeed it seems clear that these results point
towards the necessity of a realistic, three dimensional model of
thermonuclear ignition and flame spreading that considers all the
major physical effects including magnetic field, stellar spin,
chemical composition, and time variable accretion.


{}

\clearpage

\begin{figure}
\epsscale{0.9}
\plotone{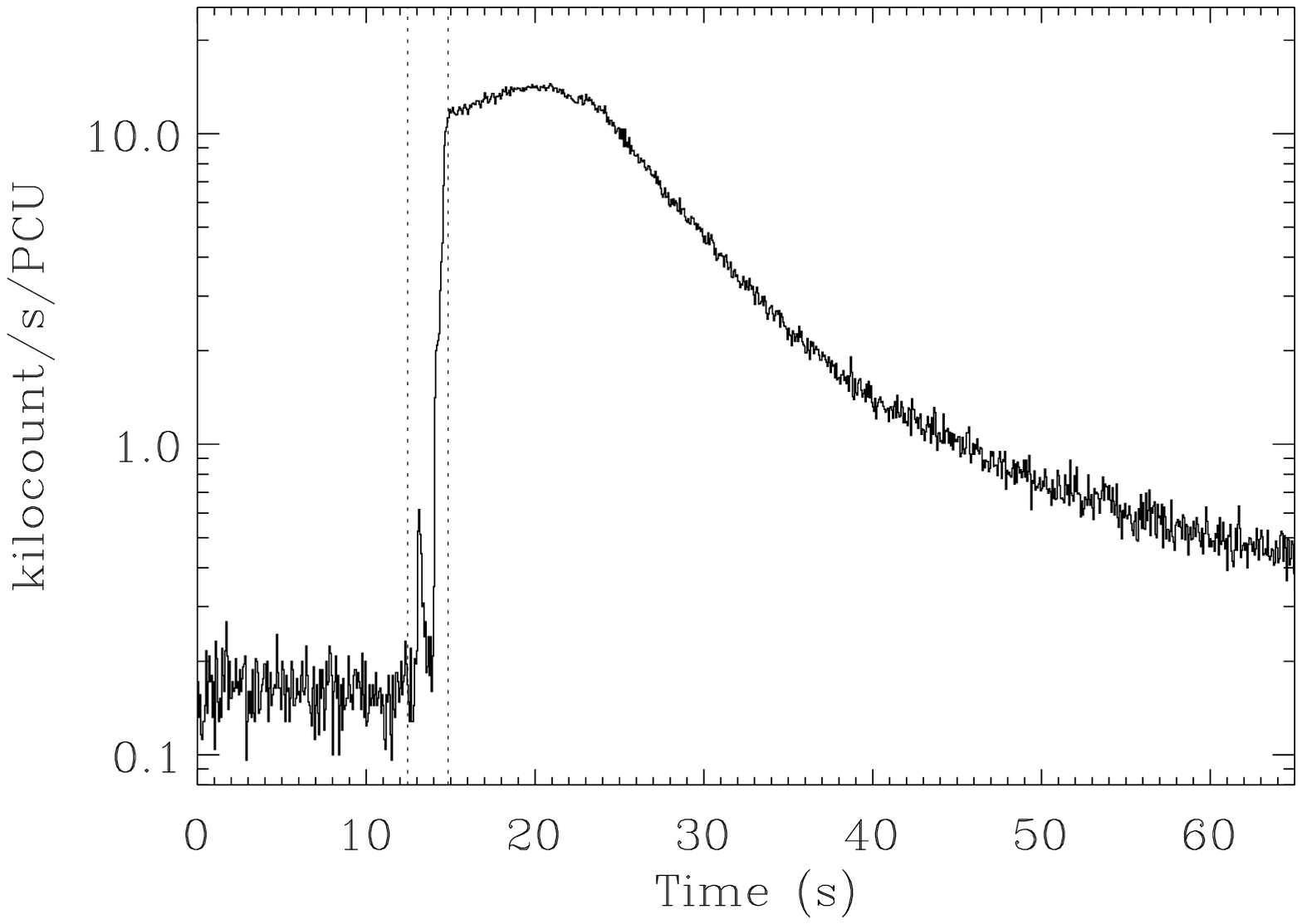}
\caption {A thermonuclear X--ray burst in the Oct. 19, 2002 {\it RXTE}
PCA data from the accreting millisecond pulsar SAX J1808.4--3658.
The burst shows a clear precursor event.  The dotted vertical lines
give the time interval that is shown in Fig. 2. Note that the vertical
scale is logarithmic.}
\end{figure}

\clearpage

\begin{figure}
\epsscale{0.8}
\plotone{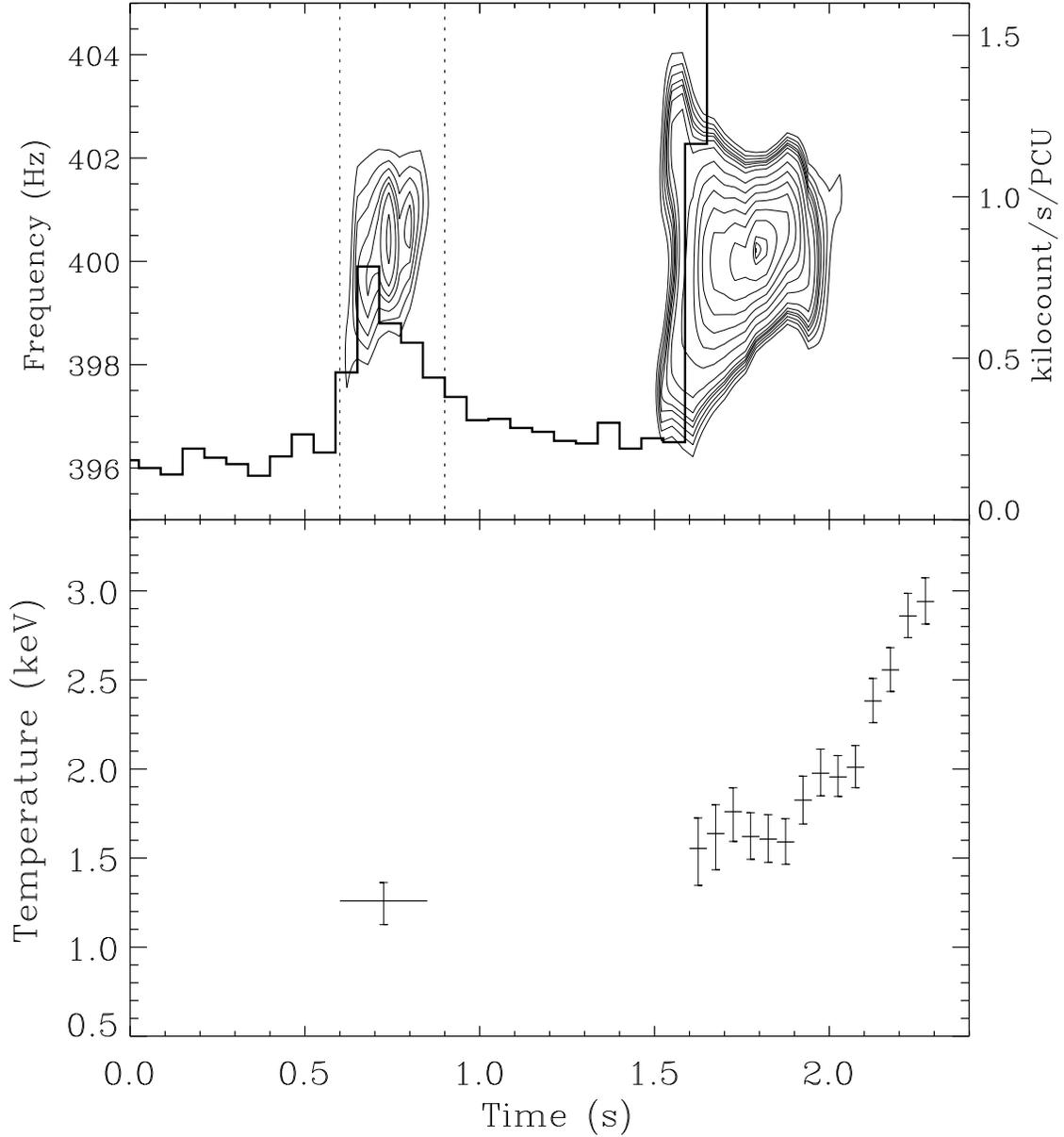}
\caption {Intensity and temperature profiles of the Oct. 19 burst.
The upper panel gives the detected intensity and the power contours
(minimum and maximum power values are $30$ and $50$ for the precursor,
and $30$ and $122.9$ for the main burst) from the dynamic power
spectra (for 0.3 s duration at 0.03 s intervals).
The dotted vertical lines give the time interval for which a power
spectrum has been shown in Fig. 3. The lower panel shows the blackbody
temperatures inferred from the model fitting of the time resolved
burst spectra (persistent emission subtracted and deadtime correction
applied).  Here the horizontal lines give the binsize and the vertical
lines give the 90\% confidence intervals.
}
\end{figure}

\clearpage

\begin{figure}
\epsscale{0.9}
\plotone{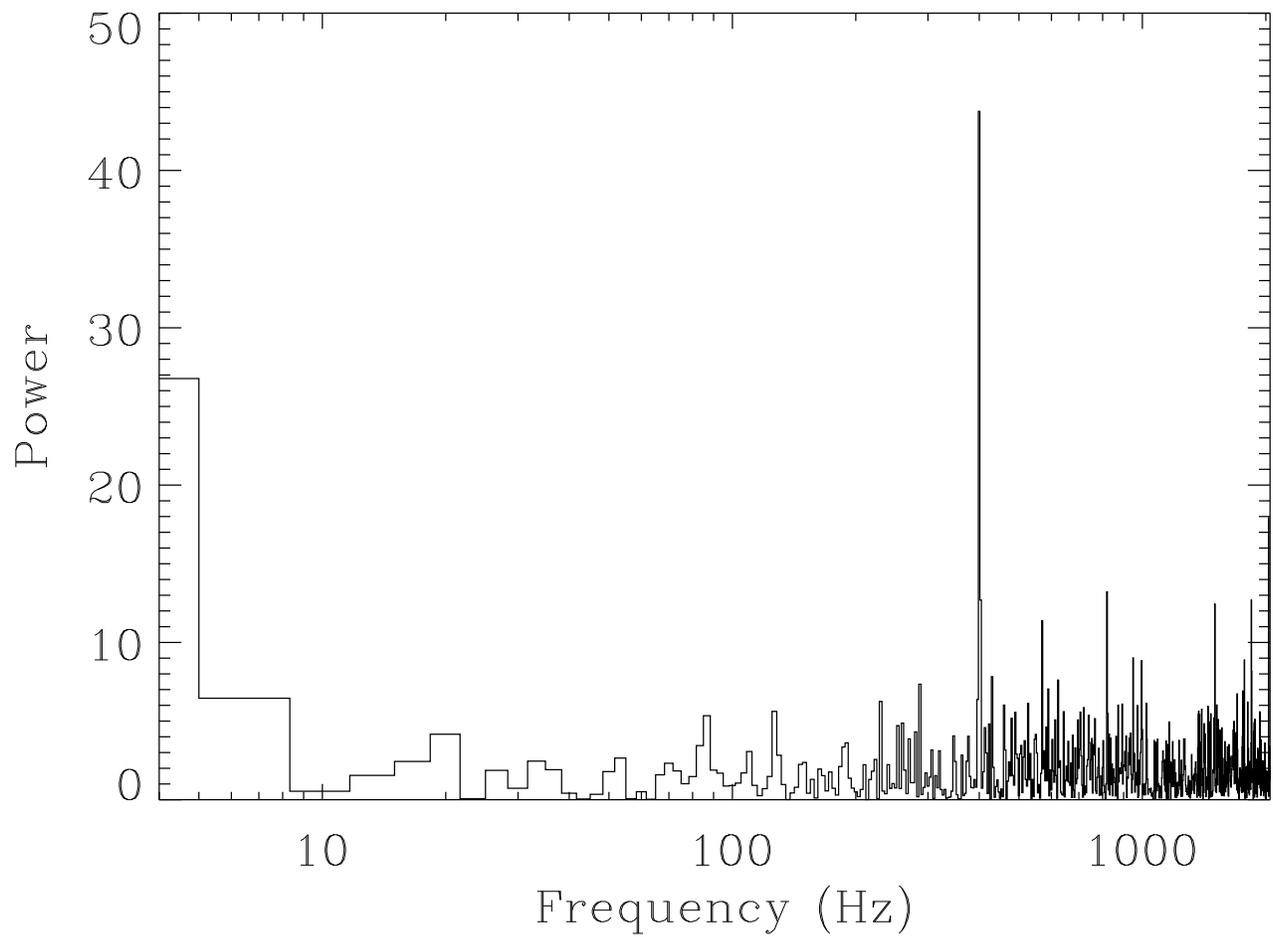}
\caption {Power spectrum for 0.3 s time interval during the
short-lived precursor burst from (see Fig. 2). The peak near 400 Hz
indicates strong burst oscillations. }
\end{figure}

\clearpage

\begin{figure}
\hspace{-3.5cm}
\epsscale{1.0}
\plotone{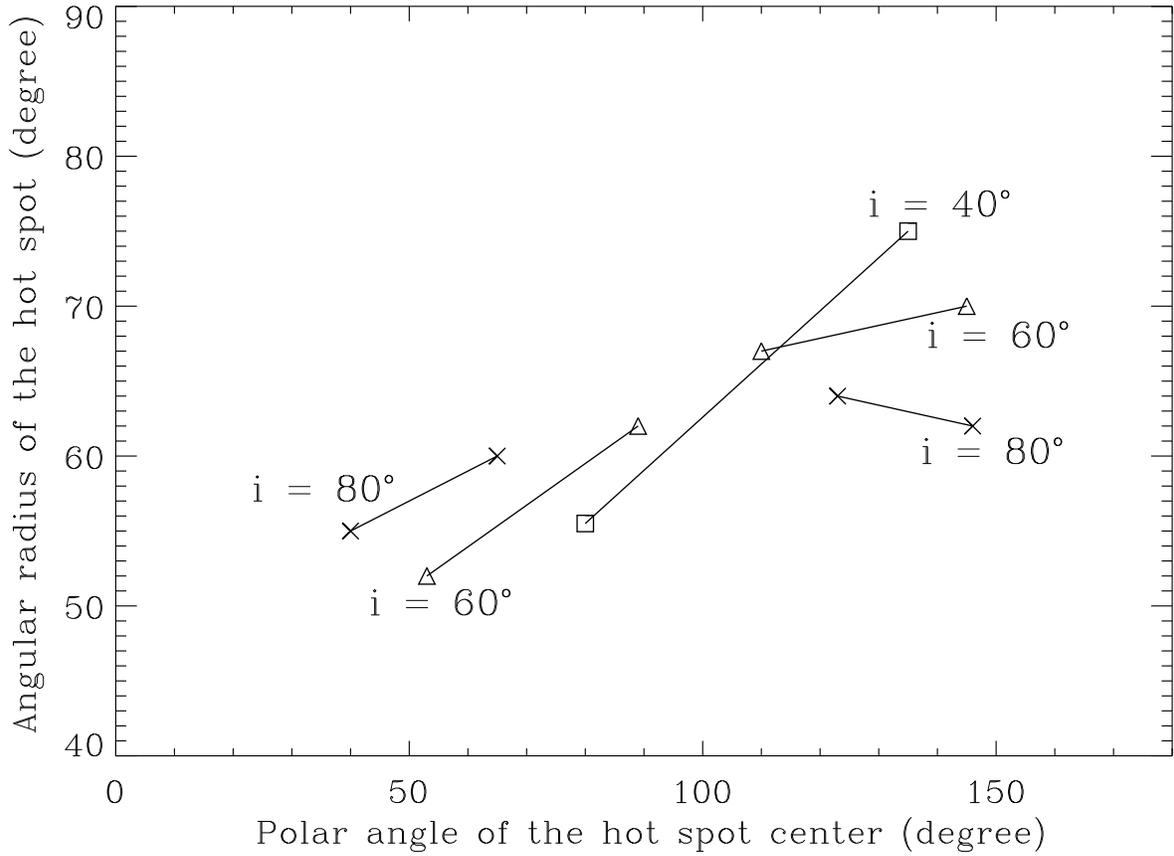}
\vspace{-8.0cm}
\caption {Allowed ranges of $\theta_{\rm c}$ and $\Delta\theta$ from
joint spectral and timing analysis. These ranges result from the observed ranges
of $Ratio$ ($=25.1\pm1.5$) and $A_2$ ($=0.403\pm0.071$; see \S~2).
Here we assume, $M=1.4$, $RC^2/GM=4.0$,
$X_{\rm main} = X_{\rm prec} = 0.7$, and three values of the observer's 
inclination angle $i$. The crosses give the $\theta_{\rm c}$ and
$\Delta\theta$ ranges for $i=80^{\rm o}$, the triangles give the ranges
for $i=60^{\rm o}$, and the squares give the range for $i=40^{\rm o}$.
Note that, for higher $i$ values, there are two disconnected allowed ranges.
}
\end{figure}

\end{document}